\documentclass[12pt,a4paper]{article}
\usepackage[T2A]{fontenc}
\usepackage[utf8]{inputenc}
\usepackage[english]{babel}
\usepackage{amssymb,graphicx}
\usepackage{amsmath,amsfonts}
\usepackage{amsthm}
\usepackage{framed}
\usepackage{fullpage}
\usepackage{feynmf}
\usepackage{microtype}
\usepackage{hyperref}

\title{Generalized Macdonald polynomials, spectral duality for
  conformal blocks and AGT correspondence in five dimensions}
\author{Yegor Zenkevich$^{a,b,c}$\thanks{yegor.zenkevich@gmail.com,
    zenkevich@ms2.inr.ac.ru}\\
  {\small $^a$\textit{ITEP, Moscow, Russia}}\\
  {\small $^b$\textit{Institute for Nuclear Research of the Russian
      Academy of
      Sciences, Moscow, Russia}}\\
  {\small $^c$\textit{NRNU Moscow Engineering Physics Institute,
      Moscow, Russia}}} \date{}

\begin{document}
\maketitle
\vspace{-55ex}
\begin{flushright}
  ITEP-TH-48/14\\
  INR-TH/2014-037
\end{flushright}
\vspace{43ex}

\begin{abstract}
  We study five dimensional AGT correspondence by means of the
  $q$-deformed beta-ensemble technique. We provide a special basis of
  states in the $q$-deformed CFT Hilbert space consisting of
  generalized Macdonald polynomials, derive the loop equations for the
  beta-ensemble and obtain the factorization formulas for the
  corresponding matrix elements. We prove the spectral duality for
  Nekrasov functions and discuss its meaning for conformal blocks. We
  also clarify the relation between topological strings and
  $q$-Liouville vertex operators.
\end{abstract}
\section{Introduction}
\label{sec:introduction}

Original Alday-Gaiotto-Tachikawa conjecture~\cite{AGT} states that
partition functions of four dimensional $\mathcal{N}=2$ supersymmetric
gauge theories in $\Omega$-background are equal to conformal blocks of
the Liouville theory. This connection has provided insight in two
dimensional CFT and gauge theory as well as other related
areas~\cite{various}. In this paper we will be concerned with a
generalization of this duality in which five dimensional
$\mathcal{N}=1$ gauge theories correspond to the $q$-deformation of
the Liouville or Toda field theories. This version of the AGT duality
has been checked for many cases in~\cite{Awata:2009ur}--\cite{AFHKSY}.

We briefly introduce the approach that we will use to study the five
dimensional AGT correspondence. As in the four dimensional case the
instanton part of the Nekrasov partition function for the $SU(2)$
gauge theory with four fundamental hypermultiplets is given by a
perturbative series in the (exponentiated) coupling constant
$\Lambda$. Each term in the expansion has in fact a finer structure
consisting of several factorized terms, so that the whole sum can be
written as a sum over pairs of partitions $(A,B)$:
\begin{equation}
  \label{eq:55}
  Z_{\mathrm{Nek}}(\Lambda| a , m_f , q, t) = \sum_{A, B} \Lambda^{|A|+|B|} \frac{(z_{\mathrm{fund}}(A))^2(z_{\mathrm{fund}}(B))^2}{z_{\mathrm{vect}}(A,B)},
\end{equation}
where $z_{\mathrm{fund}, \mathrm{vect}}$ are certain polynomials in
the gauge theory parameters which we write explicitly in
Appendix~\ref{sec:five-dimens-nekr}. On the CFT side of the AGT
correspondence this sum corresponds to the expansion of conformal
block in terms of a certain complete system of basis vectors
$|A,B,\alpha \rangle$, labelled by pairs of partitions, which can be
written schematically as
\begin{multline}
  \label{eq:35}
  \mathcal{B}\left( \Lambda \left|
  \begin{smallmatrix}
    \alpha_{\Lambda} & & \alpha_1\\
    \alpha_0 & \alpha & \alpha_{\infty}
  \end{smallmatrix}\right.
\right) = \langle V_0(0)V_{\Lambda}(\Lambda)|\text{through }
V_{\alpha} \text{ primary}| V_1(1) V_{\infty}(\infty)\rangle =\\
= \sum_{A,B} \Lambda^{|A|+|B|} \langle V_0(0)V_{\Lambda}(1)|A,B,\alpha
\rangle \langle A,B,\alpha | V_1(1)
V_{\infty}(\infty)\rangle \stackrel{\text{AGT}}{=}\\
\stackrel{\text{AGT}}{=} \sum_{A, B} \Lambda^{|A|+|B|}
\frac{(z_{\mathrm{fund}}(A))^2(z_{\mathrm{fund}}(B))^2}{z_{\mathrm{vect}}(A,B)}
\end{multline}
One way to prove the AGT correspondence is to find the special basis
for which the equality in the last two lines holds not only for each
power of $\Lambda$, but \emph{for each pair of
  partitions}~\cite{AFLT}, so that
\begin{equation}
  \label{eq:56}
  \langle
  V_0(0)V_{\Lambda}(1)|A,B,\alpha \rangle \langle A,B,\alpha | V_1(1)
  V_{\infty}(\infty)\rangle  \stackrel{\mathrm{AGT}}{=}
  \frac{(z_{\mathrm{fund}}(A))^2(z_{\mathrm{fund}}(B))^2}{z_{\mathrm{vect}}(A,B)}
\end{equation}

When $t=q$ (or $c=1$ in CFT) there is indeed a simple basis,
consisting of Schur polynomials giving the desired
expansion~\cite{AGTproof}. However, for general $t$, $q$ the basis
turns out to be a lot more elaborate: in particular the naive
deformation of Schur polynomials to Macdonald polynomials is not
enough. In this paper we show that the right basis is in fact given by
the generalization of the Macdonald polynomials $M_{AB}$ depending on
two partitions and an additional moduli
parameter~\cite{Ohkubo}. Similar polynomials for the four dimensional
case (i.e.\ generalized Jack polynomials) were introduced
in~\cite{Morozov:2013rma} and also studied in~\cite{Mironov:2013oaa}.

To compute the matrix elements in Eq.~\eqref{eq:56} we use the
$q$-deformed version of the Dotsenko-Fateev (DF) representation for
the conformal block. After the $q$-deformation the cuts in the DF
integrands pulverize into a set of poles, so that the integrals can be
taken by residues. The sum over residues is captured by the Jackson
$q$-integral, which is in fact a sum of the form
\begin{equation}
  \label{eq:42}
  \int_0^a d_q z f(z) = (1-q) \sum_{k \geq 0} q^k a f(q^k a) = \frac{1 -
    q}{1 - q^{a \partial_a}} (a f(a)).
\end{equation}

The matrix elements in this framework are given by the $q$-deformed
Selberg averages of the generalized Macdonald polynomials:
\begin{equation}
  \label{eq:57}
  \langle V_0(0)V_{\Lambda}(1)|A,B,\alpha \rangle = \frac{\int_0^1 d_q^N x
    \, \mu(x) M_{AB}(x)}{\int_0^1 d_q^N x\,
    \mu(x)},
\end{equation}
where $\mu(x)$ is certain $q$-deformed Selberg measure. To compute the
averages we devise a set of loop equation for the $q$-deformed
beta-ensemble (also called the $(q,t)$-matrix model) which provide the
recurrence relations for the averages of any symmetric polynomials.

One of our main results is the remarkable \emph{factorized} formula
for the averages of generalized Macdonald
polynomials~\eqref{eq:88}. It can be written schematically as
\begin{equation}
  \label{eq:11}
  \frac{\int_0^1 d_q^N x
    \, \mu(x) M_{AB}(x)}{\int_0^1 d_q^N x\,
    \mu(x)} =   \frac{(z_{\mathrm{fund}}(A))(z_{\mathrm{fund}}(B))}{[z_{\mathrm{vect}}(A,B)]^{1/2}}
\end{equation}
and evidently leads to the AGT conjecture~\eqref{eq:56}. Though we
were not able to obtain a rigorous proof of this formula, we have
checked it for several lower polynomials. This is the only missing
step in the proof of the five dimensional AGT conjecture, however the
conceptual picture is already apparent.

Let us clarify the relation between our study and the alternative
approach to the AGT duality proposed in~\cite{Aganagic:2013tta}. In
these works the sum over residues in the DF integrals \emph{without}
any basis decomposition was shown to be the sum over \emph{pairs of
  partitions:}
\begin{multline}
  \label{eq:76}
  \mathcal{B}\left( \Lambda \left|
  \begin{smallmatrix}
    \alpha_{\Lambda} & & \alpha_1\\
    \alpha_0 & \alpha & \alpha_{\infty}
  \end{smallmatrix}\right.
\right) = \langle V_0(0)V_{\Lambda}(\Lambda)|\text{through }
V_{\alpha} \text{
  primary}| V_1(1) V_{\infty}(\infty)\rangle \simeq\\
\simeq \int_0^1 d_q^{N_{+}} x d_q^{N_{-}} \, \mu(x) \mu(y)
\nu(\Lambda, x,y) = \sum_{R_{+}, R_{-}} \mu(q^{R_{+}} t^{\rho})
\mu(q^{R_{-}} t^{\rho}) \nu(\Lambda, q^{R_{+}} t^{\rho}, q^{R_{-}}
t^{\rho}),
\end{multline}
where $\nu(\Lambda, x, y) = \sum_{AB} \Lambda^{|A|+|B|} M^{*}_{AB}(x)
M_{AB} (y) $ is a certain rational function and $\rho =
(N-1,N-2,\ldots, 0)$ is the Weyl vector. Moreover, the integrands turn
out to combine miraculously into the Nekrasov partition function
\begin{equation}
  \label{eq:78}
  \mu(q^{R_{+}} t^{\rho}) \mu(q^{R_{-}}
  t^{\rho}) \nu(\Lambda, q^{R_{+}} t^{\rho}, q^{R_{-}} t^{\rho}) = (\Lambda^{\vee})^{|R_{+}|+|R_{-}|}
  \frac{(z_{\mathrm{fund}}(R_{+}))^2(z_{\mathrm{fund}}(R_{-}))^2}{z_{\mathrm{vect}}(R_{+},
    R_{-})}.
\end{equation}
However, this is \emph{not} the Nekrasov function featuring in the AGT
correspondence but a \emph{spectral dual}~\cite{spectral} thereof. One
can also notice that the expansion in Eq.~\eqref{eq:76} is not in the
original coupling constant $\Lambda$, which now enters each term in a
nontrivial way, but in the \emph{dual coupling} $\Lambda^{\vee}$,
i.e.\ the momentum $q^{\alpha}$ of the conformal block.

We therefore have \emph{two different} expansions of the $q$-deformed
conformal block connected by the spectral duality. The original
expansion in terms of a special basis corresponds to the AGT dual
Nekrasov function and the spectral dual expansion in the intermediate
momentum corresponds to the manifest sum over poles in the DF
integrals:
\begin{equation}
  \label{eq:85}
\parbox[c]{9.5cm}{\includegraphics[width=9.5cm]{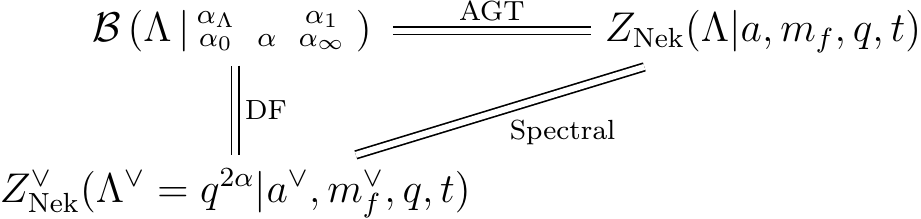}}
\end{equation}

Generally the spectral dual Nekrasov functions describe gauge theories
with different gauge groups and matter content: $SU(N)^{M-1}$ and
$SU(M)^{N-1}$ quiver theories respectively~\cite{Bao:2011rc}. In the
case at hand $N=M=2$ so the dual gauge groups are in fact the
same. Nevertheless, the parameters of the theories are reshuffled by
the duality, e.g.\ the dual coupling constant $\Lambda^{\vee}$ is a
combination of mass parameters of the original theory. Spectral
duality plays a prominent role in the Seiberg-Witten integrable
systems~\cite{IntSyst} associated with the gauge theories. One can see
that the AGT relation is a combination of the explicit DF
expansion~\cite{Aganagic:2013tta} and the spectral duality.

We can fill the missing corner in the diagram~\eqref{eq:85} by taking
the AGT dual of $Z^{\vee}_{\mathrm{Nek}}$ or interpreting
$Z_{\mathrm{Nek}}$ as a DF expansion of a dual conformal block
$\mathcal{B}^{\vee}$:
\begin{equation}
  \label{eq:4}
  \parbox[c]{9.5cm}{\includegraphics[width=9.5cm]{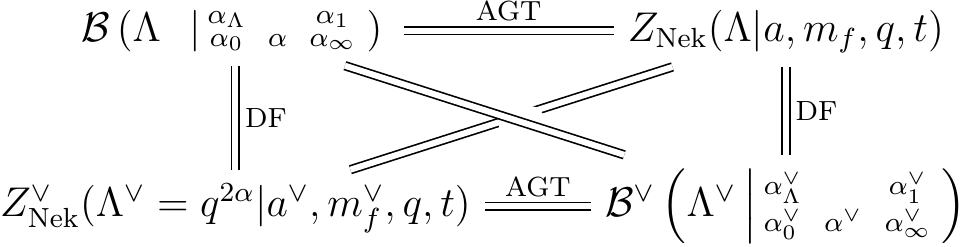}}
\end{equation}

Generally the spectral dual conformal block $\mathcal{B}^{\vee}$ has
different number of points and different conformal algebra ($q$-$W_M$
instead of $q$-Virasoro) compared to the original one
$\mathcal{B}$. Again, the case we consider here is extremely simple:
the number of points and the algebra remain the same for the
four-point conformal block of $q$-Virasoro. The dimensions of the
fields and their coordinates are however expressed through each other
in a nontrivial way. We will mention a tentative application of this
duality for conformal blocks in section~\ref{sec:conclusion}.

\begin{figure}[h]
  \begin{center}
    \begin{equation}
      \parbox[c]{6cm}{\includegraphics[width=6cm]{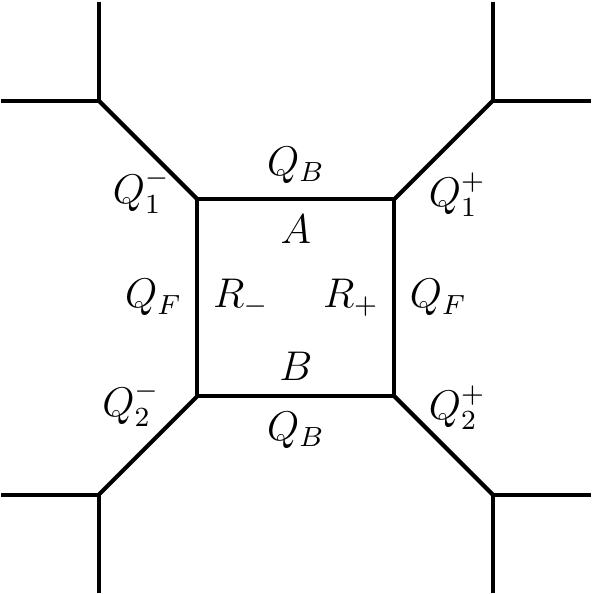}} 
      \quad = \quad
    \begin{cases}
      \sum_{A, B} Q_B^{|A|+|B|} \frac{(z_{\mathrm{fund}}(A))^2(z_{\mathrm{fund}}(B))^2}{z_{\mathrm{vect}}(A,B)} = Z_{\mathrm{Nek}}\\
      \sum_{R_{+}, R_{-}} Q_F^{|R_{+}|+|R_{-}|}
      \frac{(z_{\mathrm{fund}}(R_{+}))^2(z_{\mathrm{fund}}(R_{-}))^2}{z_{\mathrm{vect}}(R_{+},
        R_{-})} = Z^{\vee}_{\mathrm{Nek}}\notag
    \end{cases} 
    \end{equation}

    \caption{\small Toric diagram engineering an $SU(2)$ gauge theory
      with four fundamental hypermultiplets. $Q_F$ corresponds to the
      vev of the scalar, $Q^{\pm}_{1,2}$ represent the masses of the
      hypermultiplets and $Q_B \sim \Lambda$ is the coupling constant
      of the theory.}
  \end{center}
  \label{fig:1}
\end{figure}

Why do these new duality features become visible only in the five
dimensional version of the AGT correspondence and not in the original
one? It turns out the in five dimensions gauge theory partition
function has fine structure, which can be effectively analysed from
the topological string theory point of view. In the geometric
engineering approach~\cite{geom-eng}, the $\mathcal{N}=1$ gauge theory
in five dimensions is obtained by compactifying M-theory on a certain
toric Calabi-Yau threefold. Five dimensional Nekrasov partition
function is equal to the topological string partition function on the
threefold which geometrically engineers the gauge theory\footnote{The
  original geometric engineering gives the Nekrasov function in the
  self-dual $\Omega$-background, i.e.\ $t=q$. To get $t \neq q$ one
  should consider the refined topological
  strings~\cite{geom-eng-ref}.}. The geometry of this manifold is
encoded in its toric diagram. For our case of $SU(2)$ gauge theory
with four fundamental hypermultiplets the diagram is shown in
Fig.~\ref{fig:1}. The edges of the diagram correspond to the
two-cycles in the threefold while K\"ahler parameters $Q_i$ of these
cycles correspond to the gauge theory parameters.

The topological string partition function can be computed using the
topological vertex technique~\cite{Aganagic:2003db}. To each edge of
the diagram one assigns a partition and to each trivalent vertex one
assigns a certain expression $C_{\lambda \mu \nu}(q)$ depending on
three partitions residing on the adjacent edges. External edges carry
empty partitions. Partition function is obtained by summing over all
the partitions with weights $Q_i^{|\lambda_i|}$.

There are essentially two ways to carry out the sum over partitions in
Fig.~\ref{fig:1}. One can cut the diagram vertically, compute the sum
over $R_{\pm}$ explicitly and leave the sum over $A$, $B$ in the final
answer. This sum over pairs of partitions is nothing but the sum in
the Nekrasov function $Z_{\mathrm{Nek}}$. Moreover, each half of the
diagram corresponds to the matrix element in the conformal block
expansion~\eqref{eq:35}:
\begin{equation}
  \label{eq:91}
  \parbox[c]{3cm}{\includegraphics[width=3cm]{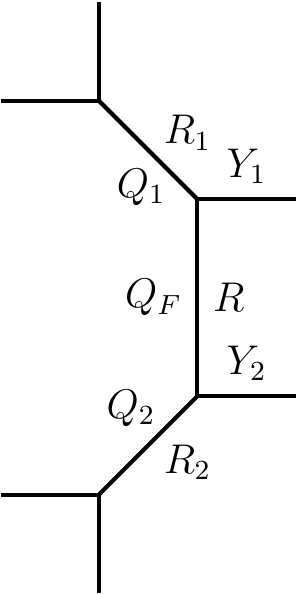}} \quad = \quad \langle V_0(0)V_{\Lambda}(1)|Y_1,Y_{2},\alpha \rangle
\end{equation}

However there is one more way to perform the summation: one can cut
the diagram \emph{horizontally} and do the sums over $A$ and $B$
first. In this way one gets the spectral dual Nekrasov function
$Z_{\mathrm{Nek}}^{\vee}$. The sum over $R_{\pm}$ corresponds to the
sum in the DF integral representation~\eqref{eq:76}:
\begin{equation}
  \label{eq:91}
  \parbox[c]{6cm}{\includegraphics[width=6cm]{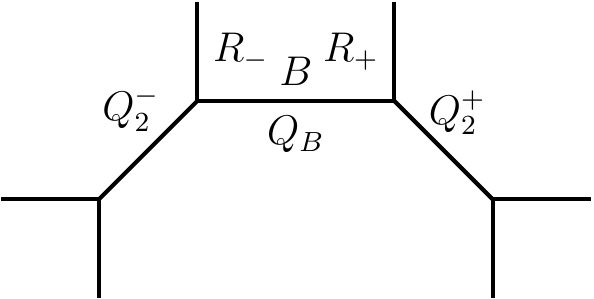}} \quad =
  \quad [\mu(q^{R_{+}} t^{\rho})
  \mu(q^{R_{-}} t^{\rho}) \nu(\Lambda, q^{R_{+}} t^{\rho}, q^{R_{-}}
  t^{\rho})]^{1/2}
\end{equation}
The spectral duality in this picture is natural and corresponds to
taking the mirror image along the diagonal (or alternatively rotating
the diagram by $\frac{\pi}{2}$).

In section~\ref{sec:gener-macd-hamilt} we introduce the generalized
Macdonald Hamiltonian, compute its eigenfunctions, i.e. generalized
Macdonald polynomials, and check their elementary properties. In
section~\ref{sec:q-deformed-dotsenko} we describe the $q$-deformation
of the Dotsenko-Fateev integral representation for conformal blocks
and introduce the decomposition in terms of generalized Macdonald
polynomials. In section~\ref{sec:loop-equations} we derive the loop
equations for the corresponding $q$-deformed $\beta$-ensemble, find
the averages of the generalized Macdonald polynomials and check the
AGT correspondence. We prove the spectral duality for $q$-deformed
Liouville correlators in section~\ref{sec:spectr-dual-gener} and
compare our results to the topological string partition function in
section~\ref{sec:comp-with-topol}. We present our conclusions in
section~\ref{sec:conclusion}.

\section{Generalized Macdonald polynomials}
\label{sec:gener-macd-hamilt}
Generalized Macdonald polynomials are symmetric polynomials in two
sets of variables $x_i$ and $\bar{x}_i$ labelled by pairs of
partitions $Y_1$, $Y_2$ and depending on an extra parameter
$Q$. Ordinary Macdonald polynomials can be obtained as the
eigenfunctions of Macdonald Hamiltonians (see
Appendix~\ref{sec:ruijs-hamilt}). Similarly, generalized Macdonald
polynomials are eigenfunctions of the following generalized MacDonald
operator, which can be thought of as the Ding-Iohara coproduct acting
on the of the original Macdonald Hamiltonian~\cite{AFHKSY}\footnote{We
  rescale $\bar{p}_n$ by $(q/t)^{n/2}$ compared to~\cite{AFHKSY}.}:
\begin{multline}
  \label{eq:22}
  H^{\mathrm{gen}}_1 = \frac{1}{t-1} \oint_{\mathcal{C}_0} \frac{dz}{z} \left[
    \left(e^{\sum_{n \geq 1} \frac{1-t^{-n}}{n} z^n p_n} e^{\sum_{n \geq 1}
      (q^n - 1) z^{-n} \frac{\partial}{\partial p_n}} - 1 \right) \right. +\\
    + \left. Q
    \left( e^{\sum_{n \geq 1} \frac{1-t^{-n}}{n} z^n ((1-t^n q^{-n})p_n + \bar{p}_n)} e^{\sum_{n \geq 1}
      (q^n - 1) z^{-n} \frac{\partial}{\partial \bar{p}_n}} - 1 \right) \right].
\end{multline}
The eigenvalues of the generalized Hamiltonian are given by the sums
of ordinary Macdonald eigenvalues
\begin{equation}
  \label{eq:37}
  H^{\mathrm{gen}}_1 M^{(q,t)}_{Y^1 Y^2} (Q|p,\bar{p}) = \left\{ \sum_{a=1}^2
    Q^{a-1} \sum_{i\geq1} (q^{Y^a_i} - 1) t^{-i} \right\} M_{Y^1 Y^2}^{(q,t)} (Q|p,\bar{p}).
\end{equation}
Eigenvalues of $H_1^{\mathrm{gen}}$ are non-degenerate, so the
eigenfunctions are uniquely determined. In practice one finds
Macdonald polynomials by solving linear equations for the coefficients
of the polynomial eigenfunctions. One writes a general symmetric
polynomial as a linear combination of $p_Y = p_{Y_1} p_{Y_2} \cdots
p_{Y_n}$, e.g. $M_Y = \sum_{Y'} c_{YY'} p_{Y'}$. Acting with
$H_1^{\mathrm{gen}}$ involves shifts of $p_n$:
\begin{equation}
  \label{eq:20}
   \exp{\left( \sum_{n \geq 1} (q^n - 1)
        z^{-n} \frac{\partial}{\partial p_n} \right)} f(p_k) = f(p_k +
    (q^k-1) z^{-k}).
\end{equation}
This gives a rational function so it is straightforward to obtain the
residues.

The properties of generalized Macdonald functions are analogous to
those of generalized Jack polynomials, though with minor
variations.

\begin{enumerate}
\item \textbf{Orthogonality, conjugation and normalization.} There is
  a convenient way to normalize generalized Macdonald polynomials:
  $M^{(q,t)}_{Y^1 Y^2} (Q |p, \bar{p}) = m_{Y^1}(p)
  m_{Y^2}(\bar{p}) + \sum_{W \neq Y} c_{YW} m_{W^1}(p)
  m_{W^2}(\bar{p})$. Conjugate Macdonald polynomials can be expressed
  through Macdonald polynomials themselves as follows
  \begin{equation}
    \label{eq:59}
    M^{*(q,t)}_{Y^1 Y^2} (Q |p_n, \bar{p}_n) = M_{Y^2
      Y^1}^{(q,t)} \left(Q^{-1} \left| \bar{p}_n, p_n - \left( 1 -
          t^n/q^n \right)\bar{p}_n\right. \right).
  \end{equation}
  Notice that the term proportional to $1-t^n/q^n$ vanishes both for
  $t=q$ and in the four dimensional limit $t=q^{\beta}$, $q \to
  1$. The polynomials form an orthogonal system:
  \begin{equation}
    \label{eq:44}
    \langle M^{*}_{A} ,
    M_{B} \rangle = \delta_{\vec{A}\vec{B}}
    \prod_{a=1,2} \frac{C_{Y^1} C_{Y^2}}{C'_{Y^1} C'_{Y^2}}
  \end{equation}
  with respect to Macdonald scalar product
  \begin{equation} \langle f(p_k) , g (p_k) \rangle = \left. f \left(
        n \frac{1 - q^n}{1 - t^n} \frac{\partial}{\partial p_k}
      \right) g(p_k) \right|_{p_k = 0}. \notag
  \end{equation}
  Here
  \begin{equation}
    \label{eq:67}
    C_A = \prod_{(i,j) \in A} (1 -
    q^{A_i - j + 1} t^{A^{\mathrm{T}}_j - i}), \qquad  C'_A =
    \prod_{(i,j) \in A} (1 -
    q^{A_i - j} t^{A^{\mathrm{T}}_j - i + 1}).
  \end{equation}

  Sometimes it is more convenient to use a different normalization
  \begin{align}
    \widetilde{M}_{Y}(p) &= G_{Y^2 Y^1}\left( Q^{-1} \right)
    \prod_{a=1,2}\prod_{(i,j) \in Y^a} (1 -
    q^{Y^a_i - j} t^{(Y^a)^{\mathrm{T}}_j - i + 1}) \widetilde{M}_Y(p)\,,\\
    \widetilde{M}^{*}_{Y}(p) &= G_{Y^1 Y^2} \left(Q \right)
    \left(\frac{t}{q}\right)^{|Y|} \prod_{a=1,2}\prod_{(i,j) \in Y^a}
    (1 - q^{-Y^a_i + j} t^{-(Y^a)^{\mathrm{T}}_j + i - 1})
    \widetilde{M}^{*}_Y(p)\,,
\end{align}
where $G_{AB}$ is given in Appendix~\ref{sec:five-dimens-nekr}. This
normalization is tailored so that the norms of the polynomials are
given by the vector part of the Nekrasov function:
\begin{equation}
\langle \widetilde{M}^{*}_{Y} (Q), \widetilde{M}_{W}(Q) \rangle =  z_{\mathrm{vect}}(Y, Q)
\delta_{Y, W}\,,
\end{equation}
Notice also that in this normalization generalized Macdonald
polynomials depend polynomially on the parameters $Q$, $q$ and $t$.
\item \textbf{The ``inversion'' relation.}
\begin{equation}
  M_Y^{(q,t)}\left( Q \left|  - \frac{1 - q^k}{1 - t^k} p_k^{(a)} \right. 
  \right) = (-1)^{|Y|} \frac{C_{Y^1} C_{Y^2}}{C'_{Y^1} C'_{Y^2}}
  M_{Y^T}^{(t^{-1}, q^{-1})} \left(
    \left. Q \right| p_k^{(a)}   \right).
\end{equation}

\item \textbf{Cauchy completeness identity.} Generalized Macdonald
  polynomials form a complete basis in the space of symmetric
  functions in two sets of variables:
  \begin{multline}
    \sum_{Y} \frac{C'_{Y^1} C'_{Y^2}}{C_{Y^1} C_{Y^2}}
    M^{*}_Y(Q|p_k^{(a)}) M_Y(Q|q_k^{(a)}) = \sum_{Y}
    \frac{\widetilde{M}^{*}_Y(Q|p_k^{(a)})
      \widetilde{M}_Y(Q|q_k^{(a)}) }{z_{\mathrm{vect}}(Y,Q)} =\\
    = \exp \left( \sum_{k \geq 1} \sum_{a=1}^2 \frac{1 - t^k}{1 - q^k}
      \frac{p_k^{(a)} q_k^{(a)}}{k} \right).\label{eq:71}
\end{multline}
\item \textbf{Specialization identities.} For $q \to 1$, $t =
  q^{\beta}$ generalized Macdonald polynomials become generalized Jack
  polynomials $M^{(q,t)}_Y(q^{2a}|p_k) \to J^{(\beta)}_Y(a|p_k)$. For
  $t=q$ the Hamiltonian $H_1^{\mathrm{gen}}$ turns into a sum of
  screening factors. Therefore, the eigenfunctions become products of
  Schur polynomials: $M^{(q,q)}_{Y^1 Y^2}(Q|p_k) = \chi_{Y^1}(p_k)
  \chi_{Y^2}(\bar{p}_k)$. This is consistent with the results
  of~\cite{AGTproof}, where Schur polynomials were found to be the
  right basis for the case of $t=q$.

  One has the following reduction to ordinary Macdonald polynomials:
  $M^{(q,t)}_{Y^1 Y^2}(Q|0 , \bar{p}_k) = M^{(q,t)}_{Y^2} (\bar{p}_k)$
  and also $M^{(q,t)}_{Y^1 \emptyset}(Q|p_k , \bar{p}_k) =
  M^{(q,t)}_{Y^1} (p_k)$. The value of generalized Macdonald
  polynomials on the Weyl vector can also be found and are nicely
  expressed through the ordinary Macdonald polynomials:
  \begin{align}
    \label{eq:79}
    M^{(q,t)}_{Y^1 Y^2} \left( Q \left|p_k = \frac{1 - Q^n}{1 - t^n},
        \bar{p}_k = - \frac{1 - \frac{t^n}{q^n}}{1 - t^n}
      \right. \right) &= M_{Y^1}^{(q,t)} \left( p_k = \frac{1 -
        Q^n}{1 - t^n} \right),\\
    M^{(q,t)}_{Y^1 Y^2} \left( Q\left|p_k = 0, \bar{p}_k = - \frac{1 -
          Q^n}{1 - t^n} \right. \right) &= M_{Y^2}^{(q,t)} \left( p_k
      = - \frac{1 - Q^n}{1 - t^n} \right).
    \end{align}

\end{enumerate}

The first few generalized polynomials are
\begin{align}
  M_{[1], [] } &= p_1, \notag\\
  M_{[],[1]} &= \bar{p}_1-\frac{p_1 (q-t)}{q \left(Q - 1\right)}\notag \\
  M_{[2],[]} &= \frac{(q+1) (t-1) p_1^2}{2 (q t-1)}+\frac{(q-1) (t+1) p_2}{2 (q t-1)},\notag \\
  M_{[1,1],[]} &= \frac{p_1^2}{2}-\frac{p_2}{2},\notag \\
  M_{[1],[1]} &= -\frac{p_1^2 (q-t) \left(q t Q+q Q-t Q-2
      t+Q\right)}{2 q \left(q Q - 1\right)
    \left(Q-t\right)}+\notag \\
  &\phantom{=} +\frac{(q-1) (t+1) Q p_2 (q-t)}{2 q \left(q Q-1\right) \left(Q-t\right)}+\bar{p}_1 p_1,\notag \\
  M_{[],[2]} &= -\frac{(q+1) (t-1) p_1^2 (q-t) \left(q^2 +q t Q-q t -t
      Q\right)}{2 q^2 \left(Q-1\right) (q t-1)
    \left(q -Q\right)}-\notag \\
  &\phantom{=}-\frac{(q-1) (t+1) p_2 (q-t) \left(q^2 -q t Q+q t -t
      Q\right)}{2 q^2 \left(Q-1\right) (q t-1) \left(q
      -Q\right)}\notag \\
  &\phantom{=}+\frac{(q+1) (t-1) \bar{p}_1 p_1 (q-t)}{(q t-1) \left(q
      -Q\right)}+\frac{(q+1) (t-1) \bar{p}_1^2}{2 (q t-1)}+\frac{(q-1)
    (t+1) \bar{p}_2}{2 (q t-1)}, \notag \\
  M_{[],[1,1]} &= \frac{p_1^2 (q-t) \left(q - t^2 Q+t Q-t \right)}{2
    q^2 \left(Q-1\right)
    \left(t Q-1\right)}-\notag \\
  &\phantom{=}-\frac{p_2 (q-t) \left(q - t^2 Q - t Q+t 1\right)}{2 q^2
    \left(Q-1\right) \left(t Q-1\right)}-\frac{\bar{p}_1 p_1 (q-t)}{q
    \left(t
      Q-1\right)}+\frac{\bar{p}_1^2}{2}-\frac{\bar{p}_2}{2}.\notag
\end{align}

\section{$q$-deformed Dotsenko-Fateev integrals}
\label{sec:q-deformed-dotsenko}
In the Dotsenko-Fateev approach conformal blocks are expressed in
terms of multiple contour integrals of the degenerate field
insertion. What is specific to the $q$-deformed case is that these
integrals can be taken by residues and reduce to multiple Jackson
$q$-integrals~\eqref{eq:42}. We use the $q$-integral formalism which
turns out to be more convenient throughout this paper. The four-point
conformal block is given by the following integral\footnote{We assume
  $v_{\pm}$, $\beta$ to be integer, though all our formulas have
  well-defined analytic continuation to non-integer values.}
\begin{multline}
  \label{eq:39}
  \mathcal{B} = \int_0^1 d_q^{N_{+}}z \int_0^{\Lambda^{-1} q/t}
  d_q^{N_{-}}z\, \prod_{i\neq j} \prod_{k=0}^{\beta-1} \left( 1 - q^k
    \frac{z_i}{z_j} \right) \prod_{i=1}^{N_{+} + N_{-}} z_i^{\alpha_0}
  \prod_{k=0}^{v_{+}-1} (1 - q^k z_i) \prod_{k=0}^{v_{-}-1} \left(1 -
    q^k \Lambda \frac{t}{q} z_i \right) =
  \\
  = C \int_0^1 d^{N_{+}}_qx \int_0^1 d^{N_{-}}_qy\, \Delta^{(q,t)}(x)
  \prod_{i=1}^{N_{+}} x_i^{u_{+}} \prod_{k=0}^{v_{+}-1} (1 - q^k x_i)\,
  \Delta^{(q,t)}(y) \prod_{i=1}^{N_{-}} y_i^{u_{-}}
  \prod_{k=0}^{v_{-}-1} (1 - q^k y_i)\times\\
  \times \prod_{i=1}^{N_{+}} \prod_{j=1}^{N_{-}} \prod_{k=0}^{\beta-1}
  \left( 1 - q^k \Lambda \frac{t}{q} \frac{x_i}{y_j} \right) \left(1 -
    q^k \Lambda \frac{x_i}{y_j}\right) \prod_{j=1}^{N_{-}}
  \prod_{n=0}^{v_{+}-1} \left( 1 - \Lambda q^{-n} \frac{t}{q}
    \frac{1}{y_j} \right) \prod_{i=1}^{N_{+}} \prod_{l=0}^{v_{-}-1}
  \left(1 - \Lambda q^l \frac{t}{q} x_i \right),
\end{multline}
where $\Delta^{(q,t)} (x) = \prod_{k=0}^{\beta - 1} \prod_{i \neq j}
(x_i - q^k x_j)$, $u_{+} = \alpha_0 + \beta (1 - N_{+} - N_{-})$ and
$u_{-} = \alpha_0 + v_{+} + \beta (1 + N_{+} - N_{-})$.

The completeness~\eqref{eq:71} of the generalized Macdonald
polynomials can be employed in the last line of Eq.~\eqref{eq:39} and
gives the following expansion
\begin{multline}
  \label{eq:3}
  \prod_{i=1}^{N_{+}} \prod_{j=1}^{N_{-}} \prod_{k=0}^{\beta-1} \left(
    1 - q^k \Lambda \frac{t}{q} \frac{x_i}{y_j} \right) \left(1 - q^k
    \Lambda \frac{x_i}{y_j}\right) \prod_{j=1}^{N_{-}}
  \prod_{n=0}^{v_{+}-1} \left( 1 - \Lambda q^{-n} \frac{t}{q}
    \frac{1}{y_j} \right) \prod_{i=1}^{N_{+}} \prod_{l=0}^{v_{-}-1}
  \left(1 - \Lambda q^l \frac{t}{q} x_i \right)=\\
  =\exp \left\{ \sum_{n \geq 1} \frac{ \Lambda^n}{n}
    \frac{1-t^n}{1-q^n} \left[ \left( -p_n \frac{t^n}{q^n} - \frac{1 -
          q^{-n v_{+}}}{1 - t^{-n}} \right) q_{-n} + p_n \left( -
        q_{-n} - \left( \frac{t}{q} \right)^n \right) \frac{1 - q^{n
          v_{-}}}{1 - t^n}  \right] \right\}=\\
  = \sum_{A,B} \Lambda^{|A|+|B|} \frac{C_A C_B}{C'_A C'_B}
  M^{*(q,t)}_{AB} \left(q^{2a} \left| - \frac{t^n}{q^n} p_n -
      \frac{1 - q^{-n v_{+}}}{1 - t^{-n}} , p_n \right. \right)\times\\
  \times M^{(q,t)}_{AB} \left(q^{2a} \left| q_{-n} , - q_{-n} - \left(
        \frac{t}{q} \right)^n \frac{1 - q^{n v_{-}}}{1 - t^n}
    \right. \right),
\end{multline}
where $p_n = \sum_{i=1}^{N_{+}} x_i^n$, $q_n = \sum_{j=1}^{N_{-}}
y_j^n$ and $C_A$, $C_A^{'}$ are given by Eq.~\eqref{eq:67}. To check
the AGT correspondence~\eqref{eq:56} one should
therefore check that
\begin{multline}
  \frac{C_{A^1} C_{A^2}}{C'_{A^1} C'_{A^2}} \left\langle
    M^{*(q,t)}_{A^1 A^2} \left(q^{2a} \left| - \frac{t^n}{q^n}
        p_n - \frac{1 - q^{-n v_{+}}}{1 - t^{-n}} ,p_n
      \right. \right)\right\rangle_{+} \times\\
  \times \left\langle M_{A^1 A^2}^{(q,t)} \left(q^{2a} \left| q_{-n} ,
        - q_{-n} - \frac{t^n}{q^n} \frac{1 - q^{nv_{-}}}{1 -
          t^n} \right. \right) \right\rangle_{-} \stackrel{\mathrm{AGT}}{=}\\
  \stackrel{\mathrm{AGT}}{=} \frac{\prod_{i=1}^{2}\prod_{f=1}^{2}
    f_{A^i}^{+} (m_f^{+} + a_i) f_{A^i}^{-} (m_f^{-} + a_i)
  }{z_{\mathrm{vect}}(A,a) } 
  \label{eq:61}
\end{multline}
where
\begin{equation}
  \label{eq:66}
\langle f(x) \rangle_{\pm} = \frac{\int d^{N_{\pm}}_qx\,
\Delta^{(q,t)}(x) \prod_{i=1}^{N_{\pm}} x_i^{u_{\pm}}
\prod_{k=0}^{v_{\pm}-1} (q^k x_i - 1) f(x)}{\int d^{N_{\pm}}_qx\,
\Delta^{(q,t)}(x) \prod_{i=1}^{N_{\pm}} x_i^{u_{\pm}}
\prod_{k=0}^{v_{\pm}-1} (q^k x_i - 1)}
\end{equation}
and the parameters of the Selberg sum are identified with the gauge
theory parameters with the help of Eqs.~\eqref{eq:54}. In the next
section we develop the loop equations for the $q$-deformed
beta-ensemble~\eqref{eq:66} in order to check Eq.~\eqref{eq:61}.

\section{Loop equations for $q$-deformed beta-ensemble}
\label{sec:loop-equations}

Let us write down the loop equations for the DF integral. They provide
the recurrence relations for the $q$-deformed beta-ensemble averages
of $p_Y = \prod_i p_{Y_i}$ and therefore determine the average of any
symmetric function.

One first observes that the $q$-integral of a total $q$-derivative is
zero
\begin{equation}
  \label{eq:41}
  \int_0^1 d_q z \frac{1}{z} (1 - q^{z \partial_z}) g(z) = 0
\end{equation}
as long as $g(1) = 0$. The loop equation is obtained from the simple
identity
\begin{equation}
  \label{eq:46}
  \int d^N_q x \sum_{i=1}^N \frac{1}{x_i} (q^{x_i \partial_i} - 1) x_i \left[ \frac{x_i
      - q}{z
      - x_i} \prod_{j \neq i} \frac{x_i - t x_j}{x_i - x_j}
    \prod_{k=1}^N \left( x_k^u \prod_{a=0}^{v-1} (q^a x_k - 1) \right)
    \Delta^{q,t}(x) f(x) \right] = 0
\end{equation}
where $f(x)$ denotes a symmetric polynomial in $x_i$ corresponding to
the insertion of extra vertex operators.  Eq.~\eqref{eq:46} in this
case is valid as a power series in negative powers\footnote{Positive
  powers give rise to total derivatives with nonvanishing boundary
  values.} of $z$. The $q$-derivative of the $q$-Vandermonde is given
by
\begin{equation}
  \label{eq:43}
  q^{x_i \partial_{x_i}} \Delta^{q,t}(x) = \left[ \left( \frac{t}{q}
  \right)^{N-1} \prod_{j \neq i} \frac{(q x_i - x_j) (t x_i -
    x_j)}{(x_i - x_j) \left(x_i - \frac{t}{q} x_j\right)} \right]
  \Delta^{q,t}(x).
\end{equation}
Using this expression
one can rewrite Eq.~\eqref{eq:46} as follows:
\begin{equation}
  \label{eq:47}
  \left\langle \sum_{i=1}^N \left[ \frac{t^{N-1} q^{u+1} (q^v x_i - 1)
      \left( q^{x_i \partial_i}f(x) \right)}{z - q
      x_i} \prod_{j \neq i} \frac{t x_i - x_j}{x_i - x_j} - \frac{(x_i
      - q) f(x)}{q (z -
      x_i)} \prod_{j \neq i} \frac{x_i - t x_j}{x_i - x_j} \right]
  \right\rangle = 0,
\end{equation}
where $\langle \ldots \rangle$ denotes the $q$-Selberg average. Let us
cast this equation into a more convenient form. We first observe that
the sum in Eq.~\eqref{eq:47} can be written as a contour integral
\begin{multline}
  \label{eq:48}
  \Biggl\langle \oint_{\mathcal{C}_x} \frac{d\xi}{\xi}
  \Biggl[\frac{t^{N-1} q^{u+1} (q^v \xi - 1) e^{\sum_{n > 0} (q^n - 1)
        \xi^n \frac{\partial}{\partial p_n} }f(p_n)}{z - q
      \xi} \prod_{j = 1}^N \frac{t \xi - x_j}{\xi - x_j}+\\
    + \frac{(\xi - q) f(p_n)}{q (z - \xi)} \prod_{j = 1}^N \frac{\xi -
      t x_j}{\xi - x_j}\Biggr] \Biggr\rangle = 0,
\end{multline}
where the contour $\mathcal{C}_x$ encircles all the points
$x_i$. Deforming the contour we pick up the residues at $0$, $z$,
$z/q$ and $\infty$. One can show that the residue at infinity contains
only positive powers of $z$ and will not affect the recurrence
relations for $p_n$, so this term can be safely dropped. Other
residues give
\begin{align}
  \label{eq:49}
  \xi = 0: &\quad t^N \left( \frac{q^{u+1}}{t} + 1 \right)
  \langle f(p_n) \rangle + \notag\\
  \xi = z: &\quad \left( \frac{z}{q} - 1 \right) \left\langle f(p_n)
    \exp \left[ \sum_{n > 0} \frac{1 - t^n}{n} z^{-n} p_n \right] \right\rangle +\notag\\
  \xi = z/q: &\quad t^{2N-1} q^{u+1} (q^{v-1} z - 1) \left\langle
    f(p_n + (1 - q^{-n}) z^n ) \exp \left[ \sum_{n > 0} \frac{1 -
        t^{-n}}{n} q^n z^{-n} p_n \right] \right\rangle = 0.
\end{align}

Analogously, assuming that $f(x) = f(p_{-n})$ is a symmetric
polynomial in \emph{negative} powers of $x_i$ one expands
Eq.~\eqref{eq:46} in \emph{positive} powers of $z$ and obtains the
recurrence relations for the negative power sums $p_{-Y} = \prod_{i}
p_{-Y_i}$:
\begin{multline}
  \label{eq:83}
  \Biggl\langle - \frac{1}{q} \left( t^{2N-1} q^{u+v+1}+1 \right)
  f(p_{-n}) + t^N \left( \frac{1}{q} - \frac{1}{z} \right) f(p_{-n})
  \exp \left( \sum_{n \geq 1} \frac{z^n}{n} (1 - t^{-n}) p_{-n}
  \right) +
  \\
  + t^{N-1} q^{u+1} \left( q^{v-1} - \frac{1}{z} \right) f(p_{-n}
    + (1 - q^n) z^{-n}) \exp \left( \sum_{n \geq 1} \frac{1}{n} z^n
      q^{-n} (1 - t^n) p_{-n} \right) \Biggr\rangle = 0.
\end{multline}

Let us demonstrate how the recurrence relations work for the simplest
example. Considering $f(p_n) = 1$ and expanding Eq.~\eqref{eq:49} to
zeroth order in $z^{-1}$ one obtains the average of $p_1$:
\begin{equation}
  \label{eq:52}
  \langle p_1 \rangle = \frac{q \left(t^N-1\right) \left(t^{N-1} q^{u+1}
      - 1\right)}{(t-1) \left(t^{2 N - 2}
   q^{u+v+2} - 1\right)}.
\end{equation}
This agrees (for $Y=[1]$) with the expression for the average of
Macdonald polynomials~\cite{AGTproof}\footnote{The overall scale
  differs by $q^{|Y|(1-v)}$. We have checked this identity up to the
  fourth level.}:
\begin{align}
  \label{eq:53}
  \langle M_Y(p_n) \rangle &= \prod_{(i,j) \in Y} \frac{q t^{i-1} (1 -
    t^{N-i+1} q^{j-1}) (1 - q^{u+j} t^{N-i})}{ (1 -
    t^{Y^{\mathrm{T}}_j - i + 1} q^{Y_i - j}) (1 -
    q^{u+v+j+1} t^{2N-i-1})},\\
  \langle M_Y(p_{-n}) \rangle &= \prod_{(i,j) \in Y} \frac{t^{i-N} (1
    - t^{N-i+1} q^{j-1}) (1 - q^{u + v -j + 2} t^{N+i-2})}{q (1 -
    t^{Y^{\mathrm{T}}_j - i + 1} q^{Y_i - j}) (1 - q^{u-j+1}
    t^{i-1})}.\label{eq:90}
\end{align}
The following simple relation for the negative power sums can be
immediately deduced
\begin{equation}
  \label{eq:89}
  \langle p_{-Y} \rangle_{u,v,N} = q^{|Y|(1-v)} \langle p_{Y}
  \rangle_{-u -v-2+ 2\beta -2\beta N,v,N}.
\end{equation}
One can also check that the averages of generalized Macdonald
polynomials are given by the \emph{factorized} expressions
\begin{multline}
  \label{eq:88}
  \left\langle M^{(q,t)}_{Y^1Y^2}\left(q_{-k},-q_{-k} - \left(
        \frac{t}{q} \right)^k \frac{1-q^{kv_{-}}}{1-t^{k}} \left|
        \frac{u_1}{u_2} = q^{-u_{-}-1} t \right. \right)
  \right\rangle_{-} = \frac{1}{G_{Y^2 Y^1}\left(
      t^{- 1} q^{u_{-} +1} \right)}\times\\
  \times \prod_{(i,j)\in Y^1} \frac{q^{1 - j + u_{-} + v_{-}} t^{2i -
      2} (t^{N_{-} - i+1}q^{j-1} - 1) (t^{-N_{-} + 2 -i}q^{j - u_{-} -
      v_{-} - 2}-1)}{q^{(Y^1)_i-j}t^{(Y^1)^{\mathrm{T}}_j-i+1}
    - 1}\times \\
  \times\prod_{(i,j)\in Y^2} \frac{(-q^{v_{-} - 1}t^{i})(t^{N_{-} -
      i}q^{u_{-} + j}-1)(t^{-N_{-} - i + 2}q^{j-v_{-}
      -1}-1)}{q^{(Y^2)_i-j}t^{(Y^2)^{\mathrm{T}}_j-i+1}
    -1} \stackrel{\eqref{eq:54}}{=}\\
  \stackrel{\eqref{eq:54}}{=} (-1)^{|Y^1|}q^{\sum_{(i,j)\in Y^1} [ j-2
    + \beta] + \sum_{(i,j)\in Y^2} [2 j -3 -2a + (2 - i)\beta ]}
  \prod_{k=1}^2 \prod_{(i,j) \in Y^k} \left( 1 - q^{(Y^k)_i - j}
    t^{(Y^k)^{\mathrm{T}}_j - i +1} \right)^{-1} \times\\
  \times \frac{\prod_{f=1}^2 \prod_{k=1}^2 f_{Y^k}^{-}(m^{-}_f +
    a_k)}{G_{Y^2 Y^1} (q^{-2a})}
\end{multline}
and analogously
\begin{multline}
  \label{eq:60}
  \left\langle M^{*(q,t)}_{Y^1 Y^2}\left(- \frac{t^k}{q^k} p_k -
      \frac{1-q^{-kv_{+}}}{1-t^{-k}}, p_k\right) \right\rangle_{+} =
  \frac{1}{G_{Y^1 Y^2}\left(
      t^{-2N_{+} + 1} q^{-u_{+} - v_{+} -1} \right)}\times\\
  \times \prod_{(i,j)\in Y^1} \frac{(-q^{-3-u_{+} - 2v_{+} +
      2j}t^{-2N_{+} - i+3})(t^{N_{+} + i-1}q^{v_{+} -j+1} -
    1)(t^{N_{+} + i-2}q^{u_{+} +v_{+}
      -j+2}-1)}{q^{(Y^1)_i-j}t^{(Y^1)^{\mathrm{T}}_j-i+1}
    -1}\times \\
  \times\prod_{(i,j)\in Y^2} \frac{q^{-u_{+} - v_{+} -j}t^{-2N_{+} +
      2i}(t^{N_{+} -i}q^{u_{+} +j} -
    1)(t^{N_{+} - i+1}q^{j-1}-1)}{q^{(Y^2)_i-j}t^{(Y^2)^{\mathrm{T}}_j-i+1} -1}\stackrel{\eqref{eq:54}}{=}\\
  \stackrel{\eqref{eq:54}}{=} (-1)^{|Y^2|}q^{\sum_{(i,j)\in Y^1} \beta
    i + \sum_{(i,j)\in Y^2}
    (2 \beta i - \beta + 2 a-j+1)}\times\\
  \times \prod_{k=1}^2 \prod_{(i,j) \in Y^k} \left( 1 - q^{(Y^k)_i -
      j} t^{(Y^k)^{\mathrm{T}}_j - i +1} \right)^{-1}
  \frac{\prod_{f=1}^2 \prod_{k=1}^2 f_{Y^k}^{+}(m^{+}_f + a_k)}{G_{Y^1
      Y^2} (q^{2a})}
\end{multline}
Combining Eqs.~\eqref{eq:88},~\eqref{eq:60} we obtain the desired
result~\eqref{eq:61}, though with additional factor corresponding to
the renormalization of the coupling constant $\Lambda \to \left( \frac{q}{t} \right)^2 \Lambda$:
\begin{multline}
\label{eq:1}
  \frac{C_{A^1} C_{A^2}}{C'_{A^1} C'_{A^2}} \left\langle
    M^{*(q,t)}_{A^1 A^2} \left(q^a , q^{-a}\left| p_n , -
        \frac{t^n}{q^n} p_n - \frac{1 - q^{-nv_{+}}}{1 - t^n}
      \right. \right)\right\rangle_{+} \times\\
  \times \left\langle M^{(q,t)}_{A^1 A^2} \left(q^a , q^{-a} \left| -
        q_{-n} - \frac{1 - q^{-nv_{-}}}{1 - t^{-n}} , q_{-n}
      \right. \right) \right\rangle_{-} =\\
  = \left( \frac{q}{t} \right)^{2(|A^1|+|A^2|)}
  \frac{\prod_{i=1}^{2}\prod_{f=1}^{2} f_{A^i}^{+} (m_f^{+} + a_i)
    f_{A^i}^{-} (m_f^{-} + a_i) }{z_{\mathrm{vect}}(A,a) }.
\end{multline}
This relation gives the proof of the five dimensional AGT conjecture
for general $q \neq t$. However, one should notice that it relies on
the formulas~\eqref{eq:88},~\eqref{eq:60} for the $q$-Selberg
averages. In Appendix~\ref{sec:loop-equat-ruijs} we sketch some ideas
which might lead to the proof of these identities. We have performed
computerized checks of Eqs.~\eqref{eq:88} and~\eqref{eq:60} on the
first three levels.

\section{Spectral duality for conformal blocks}
\label{sec:spectr-dual-gener}

In this section we demonstrate spectral duality for $q$-deformed
conformal blocks by employing the ideas of~\cite{Aganagic:2013tta}. We
write the DF integrals as explicit sums over partitions and use
combinatorial identities to cast them into the form of Nekrasov
function.

Let us examine the $q$-Selberg integral featuring in the DF
representation of the $q$-deformed CFT~\eqref{eq:66}
\begin{equation}
  \label{eq:75}
  \int d_q^N x\, \Delta^{(q,t)}(x) f(x) = \sum_{k \in \mathbb{N}^N} \Delta^{(q,t)}(x)
  f(x)|_{x_i = q^{k_i}}. 
\end{equation}
When the $q$-Vandermonde $\Delta^{(q,t)}$ is taken into account one
realizes that $x_i$ should have the form $q^{R_i} t^{N - i}$ for some
partition $R$. We therefore write the sum in the DF representation of
the conformal block~\eqref{eq:39} as a sum over pairs of partitions
\begin{equation}
  \label{eq:80}
  \frac{\mathcal{B}}{B_{\emptyset \emptyset}} = \sum_{R_{+}, R_{-}} \frac{B_{R_{+} R_{-}}}{B_{\emptyset \emptyset}},
\end{equation}
where $B_{R_{+} R_{-}}$ is the integrand from Eq.~\eqref{eq:39}
evaluated at $x_i = q^{R_{+,i}} t^{N_{+} - i}$, $y_j = q^{R_{-,j}}
t^{N_{-} - j}$. We normalize the sum so that the term corresponding to
the empty diagram is the identity.

Using the identities from Appendix~\ref{sec:useful-formulas} one
immediately obtains a compact expression for
$B_{R_{+}R_{-}}/B_{\emptyset \emptyset}$ (in the computation we
multiply all $x_i$ and $y_j$ by a factor of $q$ which does not affect
the final answer):
\begin{multline}
  \label{eq:64}  
  \frac{B_{R_{+} R_{-}}}{B_{\emptyset \emptyset}} = \prod_{(i,j) \in
    R_{+}} \frac{q^{u_{+}} t^{2i-2} (1 - q^{v_{+} + j} t^{N_{+} -i})
    (1 - q^{j-1} t^{N_{+}-i+1})}{(q^{R_{+,i} - j}
    t^{R^{\mathrm{T}}_{+,j} - i +
      1} - 1) (q^{R_{+,i} - j+1} t^{R^{\mathrm{T}}_{+,j} - i} - 1)} \times\\
  \times(1 - \Lambda q^{j-2} t^{N_{+}+2-i}) (1 -
  \Lambda q^{j+v_{-}-1} t^{N_{+}-i+1}) \times \\
  \times \prod_{(i,j)\in R_{-}} \frac{q^{u_{-}} t^{2 i -2} (1 - q^{j +
      v_{-}} t^{N_{-} - i}) (1 - q^{j-1} t^{N_{-}-i+1} )}{(q^{R_{-,i}
      - j} t^{R^{\mathrm{T}}_{-,j} - i +
      1} - 1) (q^{R_{-,i} - j+1} t^{R^{\mathrm{T}}_{-,j} - i} - 1)} \times\\
  \times (1 - \Lambda q^{-j} t^{i
    - N_{-}}) (1 - \Lambda q^{-j -v_{+} -1} t^{i-N_{-}+1})\times\\
  \times \frac{1}{G_{R_{+} R_{-}} \left( \Lambda \frac{t}{q} t^{N_{+}
        - N_{-}} \right) G_{R_{+} R_{-}} \left( \Lambda t^{N_{+} -
        N_{-}} \right)} = \\
  = (\Lambda^{\vee})^{|R_{+}| + |R_{-}|} \frac{\prod_{k=\pm}
    \prod_{f=1}^2 f^{+}_{R_k} (a^{\vee}_k + m^{\vee,+}_f) f^{+}_{R_k}
    (a^{\vee}_k + m^{\vee, -}_f)}{ z_{\mathrm{vect}}(R_{+}, R_{-},
    a^{\vee}_k)},
\end{multline}
where $f^{\pm}$, $G_{AB}$ and $z_{\mathrm{vect}}$ are given in
Appendix~\ref{sec:five-dimens-nekr},
\begin{align}
  \label{eq:92}
  a^{\vee}_{+} &= -a^{\vee}_{-} =
  a^{\vee}\notag\\
  \Lambda^{\vee} &= \Lambda^{-1} q^{u_{+} + v_{+} + 1}
  t^{N_{+}+N_{-}-2} = \Lambda^{-1} q^{u_{-}+ 1}
  t^{N_{-}-N_{+}-2} = \Lambda^{-1} q^{m_2^{-} - m_2^{+}} t^{-1},\notag\\
  2a^{\vee} &= \tau -1 + \beta(N_{+} - N_{-}+1),\notag\\
  m_1^{\vee, +} &= a^{\vee} + v_{-} + 1 + \beta(N_{-}-1),\notag\\
  m_2^{\vee, +} &= a^{\vee} + \beta N_{-},\\
  m_1^{\vee, -} &= a^{\vee} - \tau + 1 + \beta N_{-} - \beta,\notag\\
  m_2^{\vee, -} &= a^{\vee} - \tau + v_{+} + 2 + \beta N_{-} -
  2\beta,\notag
\end{align}
and $\tau = \frac{\ln \Lambda}{\ln q}$. The last line of
Eq.~\eqref{eq:64} is manifestly given by the spectral dual Nekrasov
function.

Since we have proven both horizontal and vertical equalities in
Eq.~\eqref{eq:85}, the diagonal line should also be true. We therefore
obtain a (slightly indirect) proof of the spectral duality for the
Nekrasov partition functions. Eventually, this implies that all the
dualities in the diagram~\eqref{eq:4} are valid.

\section{Comparison with topological strings}
\label{sec:comp-with-topol}
Let us now briefly discuss the meaning of our computations in the
topological string theory framework. In this section we limit
ourselves to the $t=q$ case.

Firstly we notice that the vertical half of the diagram from
Fig.~\ref{fig:1} indeed corresponds to the average of the generalized
Macdonald polynomial (which for $t=q$ is just a product of Schur
functions):
\begin{multline}
  \parbox[c]{3cm}{\includegraphics[width=3cm]{diagr-crop}} =
  \prod_{k=1}^2 \prod_{(i,j) \in Y_k}\frac{ q^{2i + m_2^{-} + a_k -j}
  }{1 - q^{(Y^k)_i - j + (Y^2)^{\mathrm{T}}_j - i + 1} }
  \frac{\prod_{f=1}^2 f_{Y_k}^{-}(m_f^{-} + a_k)}{G_{Y_1 Y_2}(q^{2a})} =\\
  = (-1)^{|Y_1|} q^{3\sum_{(i,j) \in Y_1} (i-j) + 2 \sum_{(i,j) \in Y_2} (i-j)} Q_1^{|Y_1| +
    |Y_2|} \left\langle M^{*}_{Y_1 Y_2}
  \right\rangle_{-} \label{eq:72}
\end{multline}
where
\begin{gather}
  \label{eq:74}
  Q_F = q^{2a},\qquad  Q^{-}_1 = q^{-m_1^{-} - a}, \qquad Q^{-}_2 = q^{m_2^{-} - a}
\end{gather}
This equality tells us that the vertical half of the toric diagram can
be identified with the vertex operator insertion at point $z=1$ in the
conformal block as announced in Eq.~(\ref{eq:91}).

Of course, the computations are the same for the horizontal halves of
the diagram, which correspond to the DF integrands (see
Eq.~\eqref{eq:91}):
\begin{equation}
  \parbox[c]{6cm}{\includegraphics[width=5.5cm]{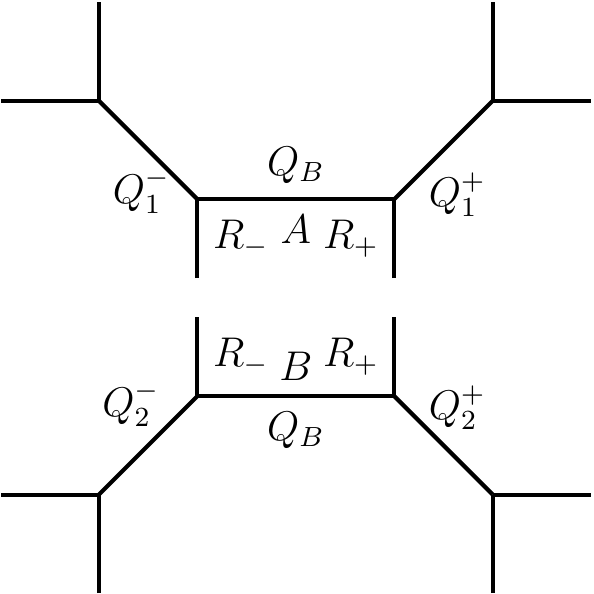}} = \frac{\prod_{k=\pm}
    \prod_{f=1}^2 f^{+}_{R_k} (a^{\vee}_k + m^{\vee,+}_f) f^{+}_{R_k}
    (a^{\vee}_k + m^{\vee, -}_f)}{ z_{\mathrm{vect}}(R_{+}, R_{-},
    a^{\vee}_k)}\notag
\end{equation}

\section{Conclusions}
\label{sec:conclusion}

In this paper we have developed techniques to study $q$-deformed
beta-ensemble and found the remarkable factorization formula for the
averages of generalized Macdonald polynomials. We have proven the
spectral duality for Nekrasov functions and conformal blocks by
investigating the explicit DF integrals. We have also found the
correspondence between the parts of the topological string partition
function and the Liouville vertex operators.

It would be interesting to better understand other features of the CFT
in the language of topological strings. In particular, the modular
properties of conformal blocks should be encoded in the topological
string partition function. The spectral duality in this case provides
an alternative expansion, which can be used to find the
nonperturbative results (cf.~\cite{Galakhov:2013jma}).

Of course extensions of the above approach to the six dimensional
gauge theories and to adjoint matter is desirable. Both of these
generalizations can be obtained in the topological string by
compactifying the toric diagram. The implications for the
Seiberg-Witten elliptic integrable systems, most interestingly to the
double elliptic ones~\cite{Dell}, should also be investigated.

\paragraph{Aknowledgements.} The author would like to thank
A.~Mironov, Al.~Morozov, And.~Morozov, S.~Shakirov and especially
S.~Mironov for stimulating discussions and criticism. The author
appreciates the hospitality of the International Institute of Physics,
Natal, Brazil where part of this work was done. The author is
supported by the Russian Science Foundation grant 14-22-00161 and by
D.~Zimin's ``Dynasty'' foundation stipend.
\label{sec:aknowledgements}

\appendix

\section{Macdonald polynomials and Ruijsenaars Hamiltonians}
\label{sec:ruijs-hamilt}
In this appendix we very briefly review some essential properties of
the (trigonometric) Ruijsenaars Hamiltonians and list some useful
expressions for them.

Macdonald polynomials are the eigenfunctions of the set of commuting
Ruijsenaars Hamiltonians $H_k$, which can be thought of as a maximal
commutative subalgebra inside the Ding-Iohara algebra~\cite{AFHKSY}:
\begin{equation}
  \label{eq:65}
   H_k M_Y = e_k(\{ (q^{Y_i}-1) t^{N-i} \}) M_Y
\end{equation}
where $e_k$ is the elementary symmetric polynomial of $N$ variables and
\begin{multline}
  \label{eq:6}
  H_k = \sum_{1 \leq i_1 < \ldots < i_k \leq N} \frac{1}{\Delta(x)}
  t^{\sum_{a} x_{i_{a}} \partial_{i_{a}}} \Delta(x)
  \left( \frac{q}{t} \right)^{\sum_{a}
    x_{i_{a}} \partial_{i_{a}}} = \\
  = t^{\frac{k(k-1)}{2}}\sum_{1 \leq i_1 < \ldots < i_k \leq N} \prod_{a = 1}^k
  \prod_{j \neq i_1...i_k} \frac{t x_{i_{a}} - x_j}{
    x_{i_{a}} - x_j} q^{\sum_{a}
    x_{i_{a}} \partial_{i_{a}}},
\end{multline}
Note that for generic $q$ and $t$ the spectrum of $H_1$ is
non-degenerate and no higher $H_k$ are needed to solve for the
eigenfunctions. However, we give some expressions for higher $H_k$ for
the sake of completeness.

Let us derive a compact expression for Ruijsenaars Hamiltonians acting
on symmetric functions in terms of the power sum variables $p_n =
\sum_{i=1}^N x_i^n$. The shift operator $q^{x_i \partial_i}$ in terms
of power sums is given by
\begin{equation}
  \label{eq:7}
  q^{x_i \partial_i} f(p_n) = f(p_n + (q^n - 1) x_i^n) = e^{\sum_{n
      \geq 1} (q^n -1)
    x_i^n \frac{\partial}{\partial p_n}} f(p_n).
\end{equation}
The shift operators commute and we get
\begin{equation}
  \label{eq:8}
  H_k= t^{\frac{k(k-1)}{2}} \sum_{1 \leq i_1 < \ldots < i_k \leq N} \prod_{a = 1}^k
  \prod_{j \neq i_1...i_k} \frac{t x_{i_{a}} - x_j}{
    x_{i_{a}} - x_j} \prod_{a=1}^k e^{\sum_{n \geq 1}(q^n -1) 
    x_{i_{a}}^n \frac{\partial}{\partial p_n}}.
\end{equation}
One can see that this sum can be expressed as a sum over
residues\footnote{We will always include the multiple of $\frac{1}{2
    \pi i }$ into the definition of the contour integral.}:
\begin{equation}
  \label{eq:9}
  H_k= \frac{t^{\frac{k(k-1)}{2}}}{(t-1)^k [k]_t!} \oint_{\mathcal{C}_x} \prod_{a=1}^{k} \frac{dz_a}{z_a}
  \prod_{a<b} \frac{z_a - z_b}{z_a- tz_b}
  \prod_{i=1}^N \prod_{a=1}^k \frac{tz_a - x_i}{z_a - x_i} e^{\sum_{a=1}^k \sum_{n \geq 1}(q^n -1) 
    z_a^n \frac{\partial}{\partial p_n}},
\end{equation}
where all $k$ integrals are taken over the same contour encircling the
points $x_i$. The origin of the $t$-deformed factorial lies in the
useful symmetrization formula
\begin{equation}
  \sum_{\sigma \in \mathfrak{S}_k} \prod_{i<j} \frac{t x_{\sigma(i)} -
    x_{\sigma(j)}}{x_{\sigma(i)} - x_{\sigma(j)}} = [k]_t!\label{eq:28}
\end{equation}
Using~\eqref{eq:28} once again for $z_a$ variables we get
\begin{equation}
  \label{eq:29}
  H_k= \frac{t^{\frac{k(k-1)}{2}}}{(t-1)^k k!}  \oint_{\mathcal{C}_x} \prod_{a=1}^{k} \frac{dz_a}{z_a}
  \prod_{a \neq b} \frac{z_a - z_b}{z_a- tz_b}
  \prod_{i=1}^N \prod_{a=1}^k \frac{tz_a - x_i}{z_a - x_i} e^{\sum_{a=1}^k \sum_{n \geq 1}(q^n -1) 
    z_a^n \frac{\partial}{\partial p_n}}.
\end{equation}
We expand the rational factors containing $x_i$ in terms of power sums
\begin{equation}
  \label{eq:10}
  H_k= \frac{t^{\frac{k(k-1)}{2} + kN}}{(t-1)^k k!}  \oint_{\mathcal{C}_x} \prod_{a=1}^{k} \frac{dz_a}{z_a}
  \prod_{a \neq b} \frac{z_a - z_b}{z_a- tz_b}
  e^{\sum_{a=1}^k \sum_{n \geq 1} \frac{1- t^{-n}}{n
    } z_a^{-n} p_n} e^{\sum_{a=1}^k \sum_{n \geq 1}(q^n -1) 
    z_a^n \frac{\partial}{\partial p_n}}.
\end{equation}

Let us now deform the integration contours so that they encircle
$z=\infty$ and $z=0$. Additional residues arise when $z_a = t z_b$.
For the first few integrals $H_k$ we have:

\begin{align}
  \label{eq:23}
  H_1 &= h_{[1]}\,,\\
  H_2 &= \frac{1}{2} h_{[1,1]} - \frac{1}{2} h_{[2]}\,,\\
  H_3 &= \frac{1}{6} h_{[1,1,1]} - \frac{1}{2} h_{[2,1]} + \frac{1}{3}
  h_{[3]}\,.
\end{align}
where\footnote{To be precise one should ensure the convergence of the
  expansions. We assume that $z_a$ are radially ordered, i.e. $|z_1| <
  \ldots < |z_k|$.}
\begin{multline} 
  \label{eq:25}
  h_{[1^k]} 
  = \frac{ t^{kN + \frac{k(k-1)}{2}}}{(t-1)^k} \left( \oint_{\mathcal{C}_{\infty}}-
    \oint_{\mathcal{C}_0} \right)^k \prod_{a=1}^{k} \frac{dz_a}{z_a}
  \prod_{a<b} \frac{(z_a - z_b)^2}{(t z_a - z_b) (z_a
    - t z_b)} \times \\
  \times   e^{ \sum_{a=1}^k (\phi_{-}(z_a) - \phi_{-}(t z_a))}
   e^{\sum_{a=1}^k (\phi_{+}(q z_a) - \phi_{+}(z_a))},
\end{multline}
\begin{equation}
  \label{eq:30}
  h_{[k]} = \frac{t^{kN}}{(t^k-1)} \left( \oint_{\mathcal{C}_{\infty}}
  - \oint_{\mathcal{C}_0}
 \right)  \frac{dz}{z} e^{\phi_{-}(z) - \phi_{-}(t^k z)} e^{\sum_{a=0}^{k-1}(\phi_{+}(q
   t^a z) - \phi_{+}( t^a z))}
\end{equation}
\begin{multline}
  \label{eq:31}
  h_{[2,1]} = \frac{t^{3N + 1}}{(t - 1)(t^2 - 1)} \left(
    \oint_{\mathcal{C}_{\infty}} - \oint_{\mathcal{C}_0} \right)^2
  \frac{dz_1}{z_1} \frac{dz_2}{z_2} \frac{(z_1 - z_2) (t z_1 -
    z_2)}{(z_1 - t z_2)(t^2 z_1 - z_2)}\times\\
\times  e^{\phi_{-}(z_1) -
    \phi_{-}(t^2 z_1) + \phi_{-}(z_2) - \phi_{-}(t z_2)}
  e^{\phi_{+}(q z_1) - \phi_{+}(z_1) + \phi_{+}(q t z_1) - \phi_{+}(t
    z_1) + \phi_{+}(q z_2) - \phi_{+}(z_2)}
\end{multline}
where $\phi_{+}(z) = \sum_{n \geq 1} z^n \frac{\partial}{\partial
  p_n}$ and $\phi_{-}(z) = \sum_{n \geq 1} \frac{z^{-n}}{n} p_n$. Let
us also note that $H_k$ can be understood as the commutative
subalgebra in the Ding-Iohara vertex algebra.

\paragraph{$t=q$ case. $GL(\infty)$ Casimir operators.}
\label{sec:cut-join-operators}
In the limit $q = t$ Macdonald polynomials degenerate into Shur
polynomials which are independent of $q$. In this limit all the
Hamiltonians can be explicitly expressed through the first one. For
example:
\begin{align}
  \label{eq:19}
  h_{[1^k]}(q) &= (h_{[1]}(q))^k,\\
  h_{[k]}(q) &= h_{[1]}(q^k),\\
  h_{[2,1]}(q) &= h_{[1]}(q^2) h_{[1]}(q).
\end{align}
The first integral also simplifies and becomes
\begin{equation}
  \label{eq:32}
  h_{[1]}(q) = \oint_{\mathcal{C}_{\infty}} \frac{dz}{z} :e^{\phi(z)}
  \frac{q^{N + z \partial_z} - 1}{q - 1} e^{-\phi(z)}:\,
\end{equation}
where $\phi(z) = \phi_{-}(z) - \phi_{+}(z)$ and the normal ordering
acts on $p_n$ and $\frac{\partial}{\partial p_n}$\footnote{One can
  fermionize the bosonic operators so that $:e^{\phi(z)}: \sim
  \bar{\psi}(z)$, $:e^{-\phi(z)}: \sim \psi(z)$. This leads to the
  fermionic construction of $GL(\infty)$ Casimirs.}.  Notice also that
$h_{[1]}(q)$ with different values of $q$ commute:
\begin{equation}
  \label{eq:33}
  [h_{[1]}(q) , h_{[1]}(q')] =0\,.
\end{equation}
The values of $h_{[1]}(q)$ are expressed through the Casimir
operators for the group $GL(\infty)$:
\begin{equation}
  \label{eq:34}
  h_{[1]}(q) \chi_R = \left( q^{N-\frac{1}{2}} C_R(q) + \frac{q^N - 1}{q - 1}\right) \chi_R\,,
\end{equation}
where $C_R(e^{\hbar}) = \sum_{i \geq 1} \left( e^{\hbar \left(R_i - i
      + \frac{1}{2}\right)} - e^{\hbar \left( - i +
      \frac{1}{2}\right)} \right) = \sum_{n \geq 0} \frac{\hbar^n}{n!}
C_R(n)$ is the generating function of the Casimirs $C_R(n) = \sum_{i
  \geq 1} \left[ \left( R_i - i + \frac{1}{2} \right)^n - (-i +
  \frac{1}{2})^n \right]$.

\section{Five dimensional Nekrasov functions and AGT relations}
\label{sec:five-dimens-nekr}
The Nekrasov partition function for the $SU(N)$ theory with $N_f = 2N$
fundamental hypermultiplets is given by
\begin{equation}
  \label{eq:5}
  Z_{\mathrm{Nek}}^{5\mathrm{d}} = \sum_{\vec{A}} \Lambda^{|\vec{A}|}
  \frac{\prod_{i=1}^{N}\prod_{f=1}^{N} f_{A_i}^{+} (m_f^{+} + a_i) f_{A_i}^{-} (m_f^{-} + a_i)
  }{z_{\mathrm{vect}}(\vec{A},\vec{a}) }\,,
\end{equation}
where $f_A^{\pm} (x) = \prod_{(i,j) \in A} \left(1 - q^{\pm x} t^{\pm
    ( i - 1)} q^{\mp (j - 1 )} \right)$,
$z_{\mathrm{vect}}(\vec{A},\vec{a}) = \prod_{i,j=1}^{N} G^{(q,t)}_{A_i
  A_j}(a_i - a_j)$ and
%
\begin{multline}
  \label{eq:2}
  G^{(q,t)}_{AB} (x)= \prod_{(i, j) \in A} \left( 1 - q^x q^{A_i - j}
    t^{B^{\mathrm{T}}_j - i + 1} \right) \prod_{(i,j) \in B}\left(1 -
    q^x q^{-B_i + j - 1} t^{-A^{\mathrm{T}}_j + i} \right) =\\
 = \prod_{(i, j) \in B} \left( 1 - q^x q^{A_i - j}
    t^{B^{\mathrm{T}}_j - i + 1} \right) \prod_{(i,j) \in A}\left(1 -
    q^x q^{-B_i + j - 1} t^{-A^{\mathrm{T}}_j + i} \right).
\end{multline}

The AGT relations for $N=2$ are:
\begin{gather}
  u_{+} = m_1^{+} - m_2^{+} - 1 + \beta\,, \qquad \qquad u_{-} = -1 +
  \beta -2a \,, \notag\\
  v_{+} = - m^{+}_1 - m^{+}_2 \,, \qquad \qquad v_{-} =
  - m_1^{-} - m_2^{-}\,, \label{eq:54}\\
  \beta n_{+} = -a + m^{+}_2\,, \qquad \qquad \beta n_{-} = a +
  m^{-}_2\,,\notag
\end{gather}
where $a_1 = - a_2 = a$. Masses $m_a$, vevs $a_i$, radius $R_5$ of the
fifth dimension and $\epsilon_{1,2}$ all have dimensions of mass. In
this paper we set the overall mass scale so that $\epsilon_1 = -b^2$,
$\epsilon_2 = 1$ and $q = e^R$. The $t$ parameter in Macdonald
polynomials is related to $q$ by $t = q^{\beta}$ with $\beta = b^2$.

\section{Ruijsenaars Hamiltonians and loop equations}
\label{sec:loop-equat-ruijs}
Let us rewrite the loop equations~\eqref{eq:47} in terms of the
Ruijsenaars Hamiltonian~\eqref{eq:6}.

We first write down a useful identity:
\begin{equation}
  \label{eq:58}
  \sum_{i=1}^N \frac{1}{z - x_i} \prod_{j \neq i} \frac{t x_i -
    x_j}{x_i - x_j}  = \frac{1}{(1 - t)z} \left[ 1 - \prod_{j=1}^N
    \frac{t z - x_j}{z - x_j} \right],
\end{equation}
which is proven by expanding the right hand side as a sum over poles
in $z$.

Using Eq.~\eqref{eq:58}, one can rewrite Eq.~\eqref{eq:47} as follows
\begin{multline}
  \label{eq:40}
  \Biggl\langle - \frac{t^{N-1} q^{u+1}}{z}H_1 f(x) + t^{N-1} q^{u+1}
  (z q^{v-1} - 1) \left[ H_1 , \sum_{n \geq 1} \frac{q^n z^{-n-1}}{q^n
      -
      1} p_n\right] f(x) + \\
  +\frac{z - q t^N}{q z (1 - t)} f(x) - \frac{z-q}{q z (1-t)}
  \prod_{j=1}^N \frac{z - t x_j}{z - x_j} f(x) \Biggr\rangle = 0,
\end{multline}
where $H_1$ is the first Ruijsenaars Hamiltonian\footnote{Curiously,
  quantum dilogarithm $\phi_q (x)$ appears in this formula: indeed
  $\sum_{n \geq 1} \frac{q^n z^{-n-1}}{q^n - 1} p_n = -\partial_z
  \sum_{i=1}^N \ln \phi_q \left( \frac{q x_i}{z} \right)$.}. The
expansion of Eq.~\eqref{eq:40} in negative powers of $z$ gives an
infinite number of $q$-Virasoro constraints which determine the
average of any symmetric polynomials in $x_i$. In fact, only two of
the infinite family of constrains are needed. Indeed, the Virasoro
generators $L_n$ with positive $n$ can be obtained by commuting $L_1$
and $L_2$. The same holds for the $q$-deformed case. We may therefore
consider only the equations obtained from the $z^{-1}$ and $z^{-2}$
terms:
\begin{gather}
  \label{eq:50}
  \Biggl\langle - t^{N-1} q^{u+v} H_1 f(x) + \frac{t^{N-1}
    q^{u+v+1}}{q-1} \left[ H_1 , p_1\right] f(x) +\frac{1 - t^N}{1 -
    t} f(x) - q^{-1}
  p_1 f(x) \Biggr\rangle = 0,\\
  \Biggl\langle \frac{t^{N-1} q^{u+2}}{(q-1)^2} \left[ H_1 , q^v p_2 -
    q (q - 1)p_1 \right] f(x) -\frac{1}{2q} ((1+t)p_2 + (1-t)p_1^2 +
  2q p_1) f(x) \Biggr\rangle = 0. \label{eq:70}
\end{gather}

For $f(x)$ a Macdonald polynomial the Hamiltonian $H_1$ acts
diagonally. However multiplication with $p_n$ produces a sum over
multiple Macdonald polynomials. To see this one should use the Pieri
idenity:
\begin{equation}
  \label{eq:68}
  e_k(p) M_Y (p) = \chi_{[1^k]}(p) M_Y(p) = M_{[1^k]}(p) M_Y(p) = \sum_W \prod_{(i,j)
    \in (C_{W\backslash Y}) \backslash (R_{W\backslash Y})}\frac{b_W(i,j)}{b_Y(i,j)} M_W(p)
\end{equation}
where the sum is over diagrams $W$ such that the skew diagram $W
\backslash Y$ is a horizontal strip (i.e. there are no more then one
box belonging to each column) of $k$ boxes; $C_{W\backslash Y}$
(resp.\ $R_{W\backslash Y}$) is the set of columns (resp.\ rows)
intersecting $W\backslash Y$ and
\begin{equation}
  \label{eq:69}
  b_Y(i,j) =
  \begin{cases}
    \frac{1 - q^{Y_i - j} t^{Y^{\mathrm{T}}_j - i + 1}}{1 - q^{Y_i - j
        + 1} t^{Y^{\mathrm{T}}_j - i}}&\text{if } (i,j)\in Y,\\
    1 &\text{else.}
  \end{cases}
\end{equation}
Using Pieri formula for $k=1, 2$ one can turn
Eqs.~\eqref{eq:50},~\eqref{eq:70} into a recurrence relations
completely in terms of the averages of Macdonald polynomials. We are
sure that one can obtain a proof of the
identities~\eqref{eq:53},~\eqref{eq:90},~\eqref{eq:88},~\eqref{eq:60}
along these lines, though we were not able to find it.

\section{Useful identities}
\label{sec:useful-formulas}
In this Appendix we list some useful combinatorial identities related
to Nekrasov functions.
\begin{gather}
  \prod_{i=1}^N \prod_{k=0}^{v-1} \frac{1 - Q q^{k + R_i} t^{-i}}{1 -
    Q q^k t^{-i}} = \prod_{(i,j)\in R} \frac{1 - Q q^{v + j-1}
    t^{-i}}{1
    - Q q^{j-1} t^{- i}} \notag\\
  \frac{\Delta^{(q,t)}(q^{R_i}t^{-i})}{\Delta^{(q,t)}(t^{-i})} =
  \prod_{k=0}^{\beta-1} \prod_{i =1}^N \prod_{j \neq i} \frac{q^{k +
      R_i} t^{-i} - q^{R_j} t^{-j}}{q^k t^{-i} - t^{-j}} =\notag\\
  = \prod_{(i,j) \in R} \frac{t^{2(i-1)} (1 - t^{N-i}q^j) (1 -
    t^{N-i+1}q^{j-1})}{(1-q^{R_i - j}t^{ R^{\mathrm{T}}_j - i+1})
    (1-q^{R_i - j +1}t^{ R^{\mathrm{T}}_j - i})} =\notag\\
  = M^{(q,t)}_R \left( \frac{1 - t^{n N}}{1 - t^n} \right)
  M^{(t,q)}_{R^{\mathrm{T}}} \left(
    \frac{1 - t^{n(1 - N)} q^{-n}}{1 - q^n} \right)\notag
\end{gather}
\begin{gather}
  \prod_{k=0}^{\beta-1} \prod_{i =1}^{N_{+}} \prod_{j = 1}^{N_{-}}
  \frac{1 - Q q^{k + R_i - P_j} t^{j-i}}{1 - Q q^k t^{j-i}} =
  \frac{\prod_{(i,j) \in R} (1 - Q q^{j-1} t^{N_{-} + 1 - i})
    \prod_{(i,j) \in P} (1 - Q q^{-j} t^{-N_{+} + i})}{G^{(q,t)}_{RP}
    (Q)} \notag\\
  \prod_{(i,j) \in R} (1- q^{R_i - j+1} t^{R^{\mathrm{T}}_j - i}) (1-
  q^{R_i - j} t^{R^{\mathrm{T}}_j - i+1}) = G_{RR}(1) \prod_{(i,j)\in
    R} (- q^j t^{i-1}) \notag\\
  G_{AB}(Q) = G_{BA}\left( \frac{q}{t Q} \right) \prod_{(i,j)\in A} (-
  Q q^{j-1} t^{1-i}) \prod_{(i,j)\in B} (- Q q^{-j} t^i) \notag
\end{gather}
where $G_{RP}$ is given by Eq.~\eqref{eq:2}.

\end{document}